\documentclass[twocolumn]{aastex63}

\usepackage{threeparttable}
\usepackage{amsmath}
\usepackage{xspace}

\newcommand*{\rom}[1]{\expandafter\@slowromancap\romannumeral #1@}

\newcommand{\msun}{M$_{\odot}$}
\newcommand{\myr}{M$_\odot$~yr$^{-1}$}

\newcommand{\ha}{H$\alpha$\xspace}

\newcommand{\nii}{[N{\sc II}]\xspace}
\newcommand{\oiii}{[O{\sc III}]\xspace}

\newcommand{\sii}{[S{\sc II}]\xspace}

\newcommand{\kms}{km\,s$^{-1}$}

\newcommand{\ergs}{erg s$^{-1}$ }

\newcommand{\htwo}{H$_{2}$}

\newcommand{\momfluxagn}{$\dot{P}_{Quasar}$ }
\newcommand{\momfluxout}{$\dot{P}_{outflow}$ }

\newcommand{\eden}{cm$^{-3}$ }

\newcommand{\momfluxratio}{$\frac{\dot{P}_{outflow}}{\dot{P}_{Quasar}}$}

\newcommand{\msigma}{$M_{\bullet}-\sigma~$}
\newcommand{\mstellar}{$M_{\bullet}-M_{*}~$}
\newcommand{\alphaco}{$\alpha_{CO}$}
\newcommand{\alphacounits}{\msun$\rm(K~kms^{-1}pc^{2})^{-1}$}

\received{May 25, 2021}
\accepted{September 28, 2021}

\submitjournal{ApJ}

\shorttitle{Multi-phase galactic outflows}
\shortauthors{Vayner et al.}

\graphicspath{{./}{figures/}}

\begin{document}

\title{Multi-phase outflows in high redshift quasar host galaxies.}

\correspondingauthor{Andrey Vayner}
\email{avayner1@jhu.edu}

\author[0000-0002-0710-3729]{Andrey Vayner}
\affiliation{Department of Physics and Astronomy, Johns Hopkins University, Bloomberg Center, 3400 N. Charles St., Baltimore, MD 21218, USA}

\author[0000-0001-6100-6869]{Nadia Zakamska}
\affiliation{Department of Physics and Astronomy, Johns Hopkins University, Bloomberg Center, 3400 N. Charles St., Baltimore, MD 21218, USA}

\author[0000-0003-1034-8054]{Shelley A. Wright}
\affiliation{Department of Physics, University of California San Diego, 
9500 Gilman Drive 
La Jolla, CA 92093 USA}
\affiliation{Center for Astrophysics \& Space Sciences, University of California San Diego,
9500 Gilman Drive 
La Jolla, CA 92093 USA}

\author[0000-0003-3498-2973]{Lee Armus}
\affiliation{Spitzer Science Center, California Institute of Technology, 1200 E. California Blvd., Pasadena, CA 91125 USA}

\author{Norman Murray}
\affiliation{Canadian Institute for Theoretical Astrophysics, University of Toronto, 60 St. George Street, Toronto, ON M5S 3H8, Canada}
\affiliation{Canada Research Chair in Theoretical Astrophysics}

\author[0000-0002-6313-6808]{Gregory Walth}
\affiliation{Observatories of the Carnegie Institution for Science
813 Santa Barbara Street
Pasadena, CA 91101
USA}

\begin{abstract}

We present Atacama Large Millimeter/submillimeter Array (ALMA) observations of six radio-loud quasar host galaxies at $z=1.4-2.3$. We combine the kpc-scale resolution ALMA observations with high spatial-resolution adaptive optics integral field spectrograph data of the ionized gas. We detect molecular gas emission in five quasar host galaxies and resolve the molecular interstellar medium using the CO (3-2) or CO (4-3) rotational transitions. Clumpy molecular outflows are detected in four quasar host galaxies and in a merger system 21 kpc away from one quasar. Between the ionized and cold-molecular gas phases, the majority of the outflowing mass is in a molecular phase, while for three out of four detected multi-phase gas outflows, the majority of the kinetic luminosity and momentum flux is in the ionized phase. Combining the energetics of the multi-phase outflows, we find that their driving mechanism is consistent with energy-conserving shocks produced by the impact of the quasar jets with the gas in the galaxy. By assessing the molecular gas mass to the dynamics of the outflows, we estimate a molecular gas depletion time scale of a few Myr. The gas outflow rates exceed the star formation rates, suggesting that quasar feedback is a major mechanism of gas depletion at the present time. The coupling efficiency between the kinetic luminosity of the outflows and the bolometric luminosity of the quasar of 0.1-1\% is consistent with theoretical predictions. Studying multi-phase gas outflows at high redshift is important for quantifying the impact of negative feedback in shaping the evolution of massive galaxies. 
\end{abstract}

\section{Introduction} \label{sec:intro}

In the local Universe, there is a noticeable dearth of baryons within massive ($M_{DM}>10^{12.5}$ \msun) galactic halos \citep{Benson03,Croton06}. Feedback from supermassive black holes (SMBH) is commonly invoked to explain the missing baryons within the massive halos. Similarly, the tight correlation between the mass of a galaxy's SMBH and the total mass of the bulge and galaxy \citep{Magorrian98,Gebhardt00} suggests that galaxies and SMBHs have evolved together. How SMBHs and galaxies co-evolve and regulate their mutual growth is an outstanding problem in modern astrophysics. Theoretical work suggests that the strong correlation between the mass of the SMBH and the velocity dispersion of the bulge (\msigma) may be achieved through quenching of star formation by powerful outflows driven once the galaxies reside on the \msigma relationship \citep{Hopkins06,Zubovas12,Zubovas14}.

The feedback that regulates the overall growth of a galaxy is expected to be most important when both the galaxies and the SMBHs were experiencing the majority of their growth \citep{Zubovas12,Choi12,Barai17,Costa18}. Given that both quasar activity and star formation peak in peak in normal galaxies at cosmic noon  ($1.5<z<3$), studying in detail the nature of outflows from quasars and starbursts at these redshifts will provide important clues about the nature of feedback during this important epoch.

Feedback can manifest itself in the form of galaxy-scale winds that expel gas from the galaxy. Moreover, winds can drive shocks and turbulence and prolong the time necessary for gas to cool, collapse and form stars. To date, most of the outflows studied in the distant ($z>1$) Universe have focused on optical emission lines that trace the warm ionized phase of outflows with typical densities $n_e\sim 100-1000$ cm$^{-3}$ and temperatures $T=10^4$ K. At the epoch of peak quasar activity $z\sim 2-3$, the bright emission lines such as \oiii\ $\lambda$5007\AA, H$\alpha$ and H$\beta$ are redshifted into near-infrared bands where they can be resolved spatially and spectrally. Using near-infrared spectroscopy, several authors have studied quasar-driven ionized winds on galaxy-wide scales both in radio-loud quasars (with powerful jets) and in radio-quiet quasars \citep{Nesvadba08,Cano-Diaz12,Carniani15,Vayner17,Vayner21,Kakkad20}.

Recent theoretical and observational works have suggested that the energy and momentum in galaxy outflows are shared between multiple phases of the gas. These gas phases span a wide range of densities and temperatures, from the dense cold molecular and neutral gas ($T\sim10-300$ K) to diffuse hot post-shock medium ($T>10^7$ K; \citealt{Crichton16,Hall19}). There have been several studies of multi-phase outflows in distant quasars, focusing on individual systems \citep{Vayner17,Brusa18,Herrera-Camus19} and at studying the cold molecular and ionized gas phase in galactic outflows. It is unknown which phase of the outflow is responsible for the bulk of the momentum and mass of these multi-phase outflows, and it may well be a function of the conditions in the host galaxies. Theoretical work suggests that the molecular gas clouds may be disrupted and entrained by the outflow \citep{Scannapieco15} or, instead, molecules may form within the outflowing gas \citep{Richings18}. Observations of nearby galaxies suggest that the cold molecular gas phase (50-100 K) may be the dynamically dominant phase of quasar winds \citep{Sun14,Cicone14,Alatalo11,Aalto12,Feruglio13,Morganti13,Veilleux17,Fiore17}. The warm (T$\sim$500 K) molecular gas phase may also be important and carry a significant fraction of the momentum in a galactic outflow \citep{Richings18}. In nearby galaxies, Spitzer observations of the warm molecular \htwo\ gas (T$\sim$500 K) have found a significant amount of gas mass in galactic outflows \citep{Beirao15,Dasyra14,Rogemar20}. 

Because cold molecular gas is the fuel for star formation, the fate of the molecular gas phase is the key link in understanding the impact of quasar feedback on star formation. Since outflows are invariably multi-phase, understanding the relationship between molecular gas and atomic gas in outflows is crucial for accurately estimating the energetics of feedback and the fate of the interstellar medium (ISM). Statistical studies are lacking, as the study of multi-gas phase outflows in the distant Universe is still in its infancy. A recent attempt at comparing molecular gas properties of galaxies with powerful (L$_{bol}=10^{45-47}$ \ergs) active galactic nuclei (AGN) to star forming galaxies with similar stellar mass have found the AGN to reside in galaxies with lower molecular gas masses hinting at potential effects of feedback from outflows or radiation \citep{Circosta21,Bischetti21}.

To address both the molecular gas reservoir and the measurement of cold molecular gas in galactic outflows requires kpc-scale spatial resolution observations that have the right sensitivity to detect the molecular gas or place stringent limits. The lowest energy transition of the \htwo\ molecule is the rotational quadrupole transitions that require gas temperatures $>100$ K to excite; hence \htwo\ is invisible in the cold molecular gas phase. The next most abundant molecule in molecular clouds is carbon monoxide (CO) that has a weak permanent dipole moment, with the near-ground rotational transitions having small excitation energies, enabling to trace colder molecular gas (5.5-55 K). In this paper, we present Atacama Large Millimeter Array (ALMA) observations of the cold molecular gas traced through rotational transitions of CO in six radio-loud quasars at $z=1.439-2.323$ with known powerful ionized gas outflows. We present observations, data reduction, and emission line analysis in Section \ref{sec:obs}. We describe how we search for molecular outflows and calculate their energetics in Section \ref{sec:dynamics}. We discuss individual objects in Section \ref{sec:indiv_obj}. We compare the entire sample, discuss potential sources that drive the multi-phase gas outflows and the dominant source of molecular gas depletion in Section \ref{sec:discussion}. We summarize our conclusions in Section \ref{sec:conc}. We use an $\rm H_{0}=67.8$ \kms\ Mpc$^{-1}$, $\Omega_{\rm m}=0.308$, $\Omega_{\Lambda}=0.692$ cosmology throughout this paper. 

\section{Observations, Data Reduction \& Line Fitting} \label{sec:obs}

\begin{deluxetable*}{ccccccccc}
\centering
\tablecaption{Summary of ALMA observations \label{tab:VLA-archive}}
\tablehead{
\colhead{Object} & 
\colhead{Date} & 
\colhead{Central frequency} &
\colhead{Continuum beam \tablenotemark{a}}&
\colhead{Continuum Sensitivity}&
\colhead{Line beam}&
\colhead{Line sensitivity}&
\colhead{Channel width} 
\\
\colhead{} & 
\colhead{} & 
\colhead{(GHz)}& 
\colhead{}&
\colhead{mJy}&
\colhead{}&
\colhead{mJy}&
\colhead{\kms}}
\startdata
7C 1354 & 2018 Jan 8-24  & 153.364 & 0.48\arcsec$\times$0.36\arcsec &0.0069&0.59\arcsec$\times$0.47\arcsec&0.08 & 34.0\\
4C 22.44 & 2017 Dec 17-30 & 135.55 & 0.39\arcsec$\times$0.35\arcsec&0.0068&0.408\arcsec$\times$0.366\arcsec&0.083 & 66.0\\
4C 05.84 & 2018 Jan 5-16  & 138.87 & 0.41\arcsec$\times$0.30\arcsec&0.0075&0.438\arcsec$\times$0.54\arcsec&0.134 & 33.0\\
4C 09.17 & 2017 Dec 25 & 	148.24 & 0.32\arcsec$\times$0.23\arcsec&0.0139&0.39\arcsec$\times$0.34\arcsec&0.130 & 15.8\\
 & 2018 Jan 8 & 	 & &&&\\
3C 318 & 2014 Jul 6  & 134.34 & 0.25\arcsec$\times$0.18\arcsec&0.0095&0.454\arcsec$\times$0.320\arcsec&0.107 & 66.66\\
3C 298 & 2016 Sep 9  & 141.85 & 0.25\arcsec$\times$0.18\arcsec&0.046&0.39\arcsec$\times$0.30\arcsec&0.22 & 34.0
\enddata
\tablenotetext{a}{From continuum images integrated over all frequencies and cleaned with robust = 0.5 parameter.}
\end{deluxetable*}

A leading goal of the ALMA observational program was to detect molecular gas outflows in quasars with powerful ionized outflows that were previously detected via optical emission line (e.g., \oiii, \ha) kinematics using integral field spectroscopy observations with adaptive optics \citep{Vayner19b,Vayner19a}. All sources selected within this study display ionized gas outflows on kpc scales, with outflow rates in the range of 50-1000 \myr, velocities $>$ 500 \kms\ with momentum fluxes $>$ 10$^{35} $ dyne and coupling efficiencies between the kinetic luminosity of the outflow and the bolometric luminosity of the quasar $>$0.05$\%$. Given the similarities between the field-of-view and angular resolution between the Keck/OSIRIS and ALMA observations we are able to study molecular gas outflows on similar spatial and dynamical time scales.

ALMA band 4 observations were conducted in Cycle 5 in the C43-5 configuration with a typical angular resolution of 0.4\arcsec\ and a maximum recoverable scale of 4-5.5\arcsec, corresponding to rough physical scales of 3 kpc and 34-43 kpc, respectively. One 1.875 GHz spectral window was tuned to the redshifted frequency of CO (3-2) or CO (4-3) emission line with an effective velocity bandwidth of 4,000 \kms, while additional three bands were tuned to the nearby continuum.\\  

Data reduction was performed using CASA (Common Astronomy Software Applications; \citealt{McMullin07}) version 5.1.2-4. The ALMA automated pipeline was used to create the measurement sets (MS) for each observing block, which were then combined into a single measurement set for each source. We performed phase self-calibration for 4C 05.84 and 7C 1354+2552, while for 3C 318 and 4C 09.17, we performed both phase and amplitude self-calibration. For each source, the band 4 quasar continuum, was used as a self-calibration model using the CASA task \textit{clean}. Given the similarity between the archival VLA images of the jets in these systems, we believe that the majority of the continuum in our sources comes from the synchrotron emission of the quasar core/jets. While there is extended continuum emission for each source, the majority of the flux is associated with the unresolved core emission from the quasar making the modeling of the continuum relatively simple for self-calibration. We image the continuum with Briggs weighting using a robust value of 0.5 and a pixel size set to 1/4th of the beam's full-width half max (FWHM). In this work, we also include our pilot observations of 3C 298 conducted in cycles 2 and 3 \citep{Vayner17} with ALMA band 4. We achieve an SNR of 600-25,000 for the peak continuum flux. The continuum SNR improved by a factor of 1.2-5 from phase-only self-calibration and in the case of 4C09.17 after amplitude self-calibration the rms further improved by a factor of 1.5, while for 3C318 we did not see a significant improvement in the rms after amplitude calibration.\\

We performed continuum subtraction using the CASA task \textit{uvcontsub} by fitting a first-order polynomial to line-free emission channels of the spectral window with the CO emission. We then subtracted the best fit continuum model from the full spectral window.\\

We imaged the cube using \textit{clean} with a robust value of 1.5 to help improve detection of fainter and more diffuse emission, resulting in a larger beam than the continuum imaging. We used a spectral pixel size of either 16 \kms, 34 \kms, or 66 \kms, depending on the signal-to-noise ratio (SNR) of the CO emission. We used a value of 0.05\arcsec\ for the spatial pixel size. For all sources except 4C 09.17, we used a wide circular aperture centered on the quasar for the cleaning mask with a radius of 1/4 the primary beam size. For 4C 09.17, the CO emission was detected in individual 16 \kms\ channels in the first cleaning cycle, and a tight mask was designed for each channel encompassing the CO (4-3) emission.

\begin{figure*}
    \centering
    \includegraphics[width=7.0in]{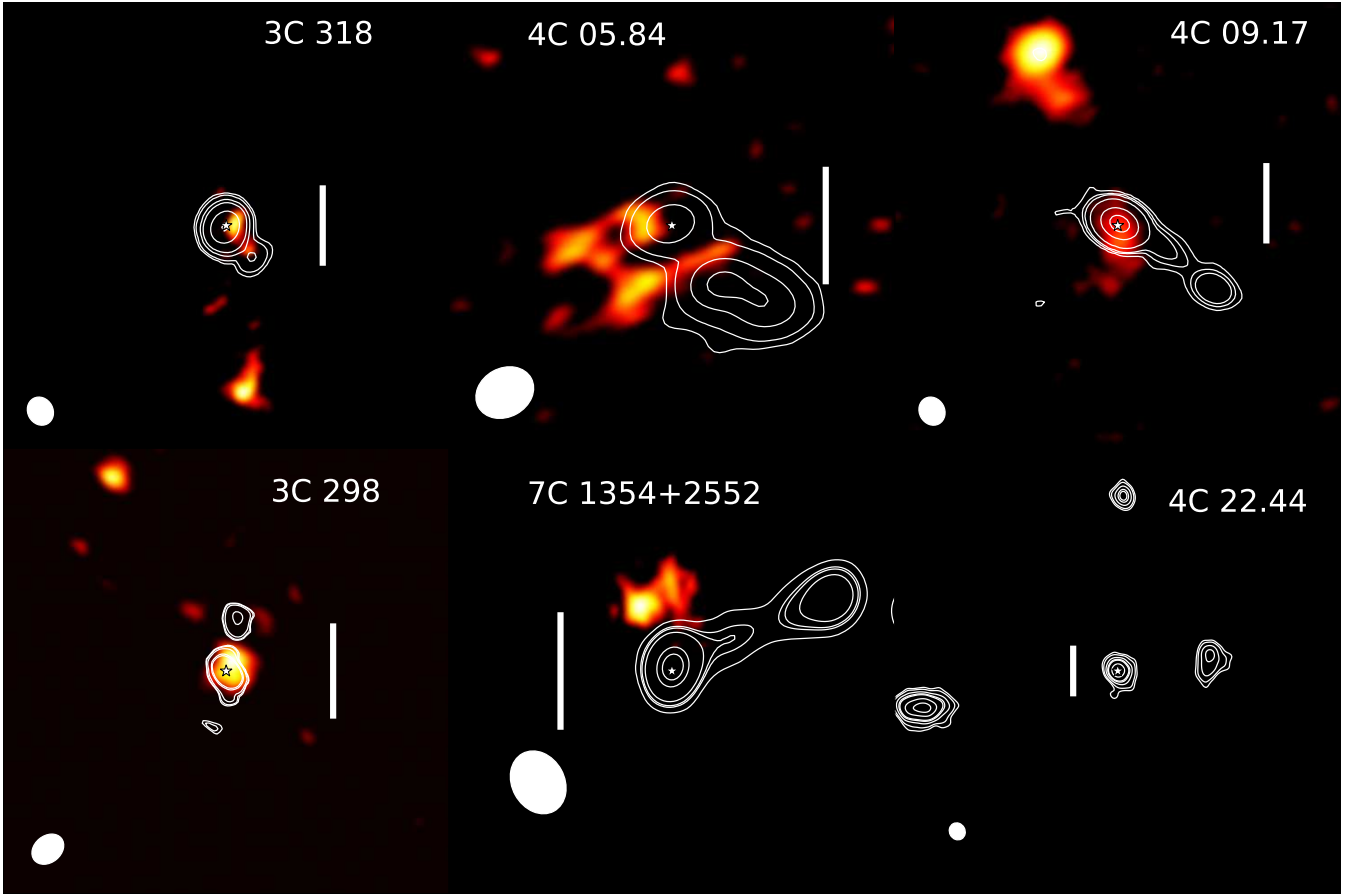}
    \caption{ALMA band 4 observations of the sources within our sample. We present ALMA band 4 integrated intensity maps of CO emission. In contours we present the ALMA band 4 continuum emission that is dominated by synchrotron emission from the quasar jets. The contours stretch from a peak flux of 0.013, 0.0037, 0.2, 0.003, 0.045, and 0.002 Jy/beam down to 2$\sigma$ sensitivity in linear steps, from top left to bottom right, respectively. The ellipse in the lower-left corner of each map represents the beam of the emission line data. The star in the center represents the quasar's location, and the bar represents 1 arcsecond. The maps are at at a position angle of 0$^{\circ}$, where north is oriented up, the only exception is 3C 298 that is at a position angle of 103$^{\circ}$ East of North.}
    \label{fig:all_sources_CO}
\end{figure*}

To search for extended emission, we construct an SNR map for each channel in the data cube. An SNR map is made by first computing a standard deviation in a large aperture away from the phase center, followed by dividing the flux per spaxel by the standard deviation. SNR maps are constructed from data cubes that were not corrected for the primary beam's response. For any spaxel containing emission with a peak SNR$\geq$4 per beam, we fit the emission line in neighboring spaxels that lie in the beam with a Gaussian plus constant continuum model using the least-squares fitting routine \textit{curvefit} within \textit{SciPy}. We create a 0$^{th}$ moment (flux) map by integrating the emission line in each spaxel with a successful emission line fit over the fitted Gaussian model and a 1$^{st}$ moment map by computing the line centroid's Doppler shift relative to the redshift of the quasar host galaxy. The integrated intensity maps are all optimally extracted, with a varying window for integrating the CO emission line based on the velocity offset and dispersion. We construct a 2$^{nd}$ moment map using the fit's velocity dispersion from the Gaussian model. All moment maps are constructed from data cubes that were corrected for the primary beam's response. Figure \ref{fig:all_sources_CO} showcases the integrated CO (3-2) or CO (4-3) maps for sources with detected extended emission.\\

Herein we use the quasar redshifts that are reported from Keck/OSIRIS observations \citep{Vayner19b} where they were derived from the centroid of the spatially unresolved ($<1.5$ kpc) \oiii\ or \ha emission lines, effectively originating from the narrow-line-region (NLR) of the quasar. The \oiii\ and \ha\ redshifts agree within the centroiding uncertainty. For 3C 298 the redshift derived from the molecular gas disk traced through CO (3-2) is within 50 \kms\ of the redshift derived from the quasar emission \citep{Vayner17}.

\begin{deluxetable*}{lcclll@{\extracolsep{-10pt}}c}[!th]
\tiny
\tablecaption{Sample properties \label{tab:sample}}
\tablehead{
\colhead{Name} & 
\colhead{RA} & 
\colhead{DEC} &
\colhead{z\tablenotemark{a}} &
\colhead{L$_{\rm bol}$} &
\colhead{L$_{\rm 178 MHz}$} &
\colhead{M$_{\rm BH}$}\\
\colhead{} & 
\colhead{J2000} &
\colhead{J2000} & 
\colhead{} & 
\colhead{($10^{46}$ \ergs)} & 
\colhead{($10^{44}$ \ergs)} & 
\colhead{\msun}}
\startdata
4C 09.17 & 04:48:21.74  & +09:50:51.46 & 2.1170 & 2.88$\pm$0.14 &2.6 & 9.11  \\
7C 1354+2552 & 	13:57:06.54 & +25:37:24.49 & 2.0068 & 2.75$\pm$0.11 & 1.4& 9.86 \\
3C 298 & 14:19:08.18 & +06:28:34.76  & 1.439\tablenotemark{b} & 7.80$\pm$0.30 & 12 &9.51 \\
3C 318 & 15:20:05.48  & +20:16:05.49 & 1.5723 & 0.79$\pm$0.04 & 4.0 &9.30 \\
4C 22.44 & 17:09:55.01 & +22:36:55.66 & 1.5492 & 0.491$\pm$0.019 & 0.6 &9.64 \\
4C 05.84 & 22:25:14.70  & +05:27:09.06 & 2.320 & 20.3$\pm$1.00& 4.5 &9.75  \\
\enddata
\tablenotetext{a}{Redshift relative to narrow-line region emission of the quasar, derived from 5007 \AA\ \oiii\ emission.}
\tablenotetext{b}{Redshift derived from host galaxy CO (3-2) emission \citep{Vayner17}}
\end{deluxetable*}

\section{Dynamics of the molecular gas}\label{sec:dynamics}

\subsection{Systemic molecular gas}

For each source, we search for emission at the systemic redshift of the quasar host galaxy. Narrow, CO line emission is found in the host galaxies of 4C 09.17A, 4C 09.17B, 3C 298, and 7C 1354+2552. The narrow CO emission in 3C 298 resembles a rotating disk that we modeled in \citet{Vayner17}. The narrow emitting gas in the other systems does not show a smooth velocity gradient that would be indicative of a rotating galactic disk. 

\noindent We compute the emission line luminosity using the following equation:

\begin{equation}
   L^{'}_{CO}=3.25 \times 10^{7}S_{CO}\Delta v \frac{D^{2}_{L}}{(1+z)^{3}\nu_{obs}^{2}} \rm K~km~s^{-1}~pc^{2},  
\end{equation} 

\noindent 
where $\nu_{obs}$ is the observed CO transition frequency, $D_{L}$ is the luminosity distance, and $S_{CO}\Delta v$ is the line integrated flux in units of Jy \kms. We convert the observed CO transition luminosity into CO (1-0) luminosity (L$^{'}_{CO(1-0)}$) by assuming that the low-J CO transitions are thermalized and are optically thick, so ${L}_{\mathrm{CO}\ 4\mbox{--}3}^{{\prime} }={L}_{\mathrm{CO}\ 3\mbox{--}2}^{{\prime} } ={L}_{\mathrm{CO}\ 1\mbox{--}0}^{{\prime} }$. Using the ratios (${r}_{{\rm{J}}1}={L}_{\mathrm{CO}J\to J-1}^{{\prime} }/{L}_{\mathrm{CO}\ 1\mbox{--}0}^{{\prime} }$) from \cite{CarillinWalter13} with $r_{31}$=0.97 and $r_{41}$=0.87 did not significantly change our results. Furthermore in 3C 298 we found that the molecular gas is consistent with being thermalized and optically thick \citep{Vayner17}. These physical conditions are consistent with what is found for Mrk 231 \citep{Feruglio15}. Finally, we convert the ${L}_{\mathrm{CO}\ 1\mbox{--}0}^{{\prime}}$ line luminosity into molecular gas mass using the CO-to-H$_{2}$ conversion factor: $\alpha_{\rm CO}$ with units of (K \kms$\rm pc^{2}$)$^{-1}$. For sources where we do not detect any narrow CO emission at the systemic redshift, we place a limit on the molecular gas mass over an aperture equal to the beam size for an emission line with a velocity FWHM of 250 \kms. The molecular gas mass limits can be linearly scaled with a different \alphaco\ value. For sources with detected molecular gas at the quasar's systemic redshift, we compute the radius of the molecular gas region, which allows us to measure the gas surface density. All radii are computed using a curve of growth method. Effective radii refer to a region which encloses 50\% of the flux, while a ``maximum" extent refers to a size scale which encloses 90\% of the flux. In all cases, narrow emission at the systemic redshift of the quasar is spatially resolved by our observations. We deconvolve the size of the beam from all radius measurements. For sources with no detected CO emission, we use the molecular gas mass limit and the beam of the observations as a proxy for the radius. Values associated with the molecular gas at the systemic redshift are summarised in Table \ref{tab:narrow-prop}.

\begin{deluxetable*}{lcccccr}[!th]
\tablecaption{Properties of molecular gas at the systemic redshift of each quasar. $S_{\nu}\Delta V$ is the spatially and line-integrated CO flux. $M_{H_{2}}$ is the mass of gas at the systemic redshift of each galaxy, assuming an $\alpha_{CO}$ of 0.8. $R$ is the radius of the region. V$\rm_{\sigma}$ is the velocity dispersion. SFR is the expected star formation rate based on the currently available molecular gas reservoir based on the KS law.  \label{tab:narrow-prop}}

\tablehead{
\colhead{Source}&
\colhead{$S_{\nu}\Delta V$}&
\colhead{$M_{H_{2}}$}&
\colhead{R}&
\colhead{V$\rm_{\sigma}$}&
\colhead{$\Sigma_{molecular}$}&
\colhead{SFR}\\
\colhead{}&
\colhead{Jy \kms}&
\colhead{$\times10^{9}$\msun}&
\colhead{kpc}&
\colhead{\kms}&
\colhead{\msun pc$^{-2}$}&
\colhead{\myr}}
\startdata
4C 09.17 A RL & 0.29$\pm$0.03 & 3$\pm$0.3 & 2.8  & 158.0$\pm$14 & 137$\pm$13 & 6 \\
4C 09.17 B RQ & 0.88$\pm$0.09 & 10$\pm$1 & 1.1 & 143.5$\pm$12 & 2357$\pm$235 & 56 \\
7C 1354 & 0.028 $\pm$0.003 & 0.3$\pm$0.1 & 2.0 & 44.10$\pm$11 & 23$\pm$2 & 1\\
3C 298 & 0.63$\pm$0.07 & 6.6$\pm$1 & 1.6 & 42.35$\pm$12.8 & 820$\pm$10 & 24\\
3C 318 & $<0.05$ & $<0.6$ & -- & -- & -- & --\\
4C 22.44 & $<$0.05 & $<$ 1 & -- & -- & -- & --\\
4C 05.84 & $<0.07$ & $<0.8$ & -- & -- & -- & --\\
\enddata
\end{deluxetable*}

\subsection{Observed Molecular Outflows}\label{sec:search_outflow}

\begin{figure*}[!th]
    \centering
    \includegraphics[width=7.0 in]{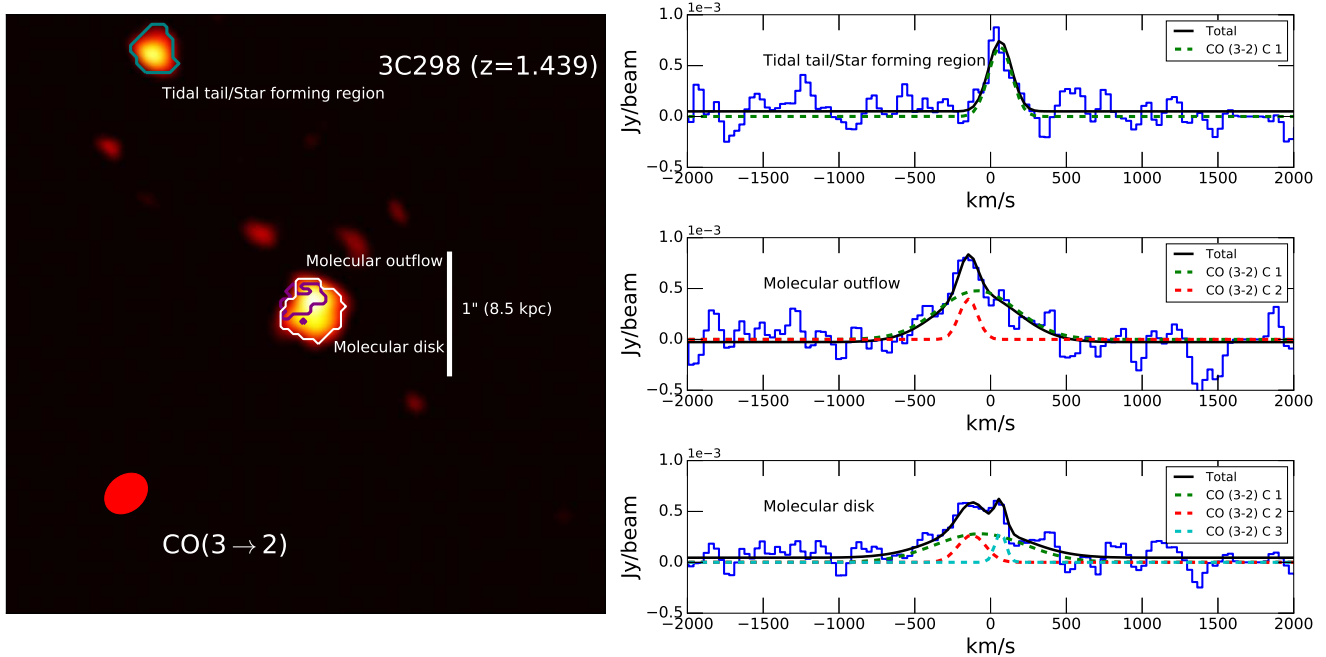}
    \caption{ALMA band 4 observations of 3C 298. On the left we show optimally extracted intensity map of CO emission in the 3C 298 system, detected in our 2017 study of this object \citep{Vayner17}. The purple contours outline the molecular outflow region, white contours outline total emission from the molecular disk, while the teal contours outline a star forming/tidal tail feature. On the right we show a spectrum integrated over each distinct region along with fit to the CO (3-2) emission line. Dashed lines represent the individual Gaussian components of the emission line fit, while the solid black line represents the sum of all components and a 0th order polynomial fit to any residual continuum. The systemic redshift of the quasar host galaxy is at 0 \kms. The ellipse in the lower left corner on the right panel represents the beam of the ALMA band 4 observations.}
    \label{fig:3C298_spec}
\end{figure*}

Ionized gas outflows typically show a peak emission line offset  $>$ 300 \kms\ from the quasar redshift measured from the narrow emission line region, with a velocity dispersion $>$ 250-300 \kms\ over the outflow region for the sources within our sample \citep{Vayner19b}. Based on these observed velocities of the ionized gas outflows, we define the criteria for a molecular outflow to be any molecular gas that has a peak emission line offset relative to the redshift of the quasar host galaxy $|v| >300$ \kms\ or a spaxel with a velocity dispersion greater than 250 \kms. The selected outflow criteria are typical of molecular outflows found in nearby galaxies \citep{Fluetsch19}. Using these criteria, we detect molecular gas outflows in four quasars in the ALMA sample. For 3C 318 and 4C 05.84, we select outflows based on both broad ($\sigma>250$ \kms) and offset ($|v| > 500$ \kms) emission and for 4C 09.17, based on broad ($\sigma>300$ \kms) emission. We include the molecular gas outflow in 3C 298 that was previously detected in \citet{Vayner17} based on broad CO (3-2) and CO (5-4) emission. For each source we extract a spectrum by integrating over all spaxels satisfying this outflow criteria. We present the extent of the molecular outflows along with the spectra in Figures \ref{fig:3C318_spec}, \ref{fig:4C0917_spec}, \ref{fig:4C0584_spec}, and \ref{fig:3C298_spec}. 

We potentially may be missing dense outflowing gas moving at slower speeds since our high velocity criteria is based on atomic gas observations, which may lead us to underestimate the total molecular gas in the outflow for the 4C 09.17 and 3C 298 systems where we detect CO emission moving at speeds $<$ 300 \kms. Observations of denser molecular gas tracers, such as CS or HCN, could provide a more complete picture of the outflow. \\

We calculate the molecular gas mass in the outflows based on flux associated with the broad or highly offset emission line component. We use an \alphaco\ value of 0.8 \alphacounits for consistency with other works at low and high redshift \citep{Herrera-Camus19,Fluetsch19}, this value is commonly adopted for the molecular gas in the ISM of nearby Ultra Luminous Infrared Galaxies \citep{Bolatto13}. However, in some well-studied molecular outflows in nearby galaxies, the conversion factor can be much higher, $\alpha_{\rm CO}\sim 2$ \citep{Cicone18}, so we may be conservative with our mass estimates.

We compute the molecular gas outflow rate using

\begin{equation}\label{equation:outflow-thin-shell}
    \dot{M}_{H_{2}}=\frac{M_{H_{2}}v_{out}}{R_{out}}.
\end{equation}

We select this equation since the molecular gas outflows in 3C 318, 4C 05.84, and 4C 09.17-A RL are seen as a single high-velocity offset component that spans either blue- or red-shifted velocities. The molecular gas outflows in 4C 09.17B and 3C 298 may be closer in geometry to a filled wide-angle cone since they span a broader velocity range. In these two sources, the estimates obtained from equation \ref{equation:outflow-thin-shell} should be multiplied by a factor of 3 if the outflows are closer to a filled cone. Here $R_{out}$ is the extent of the outflow where 90\% of the flux associated with the outflow emission accumulates. The velocity is computed as $v_{out} = \left | v_{r} \right |  + FWHM/2$, where $\left | v_{r} \right |$ and $FWHM$  are the radial velocity and full-width-half-maximum in units of \kms\ of the emission relative to the systemic redshift of the emission associated with the outflow component.

In addition to the outflow rates we also compute the momentum flux of the outflow using: 

\begin{equation}
    \dot{P}_{H_{2}} = \dot{M_{H_{2}}} \times v_{out}
\end{equation}

\noindent and the kinetic luminosity:

\begin{equation}
    \dot{E}_{H_{2}} = \frac{1}{2}\times\dot{M_{H_{2}}}\times v_{out}^{2}.
\end{equation}

The CO line luminosity used to compute the molecular gas mass along with the spatial extent, velocity, outflow rate, and energetics are summarized for each source in Table \ref{tab:outflow-prop}.

\begin{deluxetable*}{lllccccccccl}
\tablecaption{Multi-phase outflow properties. $S_{\nu}\Delta V$ is the spatially and line-integrated CO flux associated with the molecular outflow. R$\rm_{out}$ is the radial extent of the outflow. V$\rm_{out}$ is the velocity of the outflow. dM/dt$\rm_{H_{2}}$ and dM/dt$\rm_{Ionized}$ are the molecular and ionized outflow rates. $\dot{P}_{H_{2}}$ is the momentum flux of the molecular outflow (assuming \alphaco$=0.8$), while $\dot{P}_{Ionized}$ is the momentum flux of the ionized outflow, using the dM/dt$\rm_{H\alpha}$ outflow rate from \citep{Vayner19a}. $\frac{\dot{P}_{outflow}}{\dot{P}_{Quasar}}$ is the ratio of the momentum flux of outflow to the momentum flux of the quasar accretion disk ($L_{bol}/c$) using a sum of the ionized and molecular outflow momenta flux.\label{tab:outflow-prop}}

\tablehead{\colhead{Source}&
\colhead{$S_{\nu}\Delta V$}&
\colhead{M$\rm_{H_{2}}$}&
\colhead{R$\rm_{out}$}&
\colhead{V$\rm_{out}$}&
\colhead{V$\rm_{FWHM}$}&
\colhead{dM/dt$\rm_{H_{2}}$}&
\colhead{M$\rm_{Ionized}$}&
\colhead{dM/dt$\rm_{Ionized}$}&
\colhead{$\dot{P}_{H_{2}}$}&
\colhead{$\dot{P}_{Ionized}$}&
\colhead{$\frac{\dot{P}_{outflow}}{\dot{P}_{Quasar}}$}\\
\colhead{}&
\colhead{Jy \kms}&
\colhead{$\times10^{9}$\msun}&
\colhead{kpc}&
\colhead{\kms}&
\colhead{\kms}&
\colhead{\myr} &
\colhead{$\times10^{9}$\msun} &
\colhead{\myr}&
\colhead{$10^{35}$dyne}&
\colhead{$10^{35}$dyne}&
\colhead{}
}
\startdata
4C 05.84 & 0.1$\pm$0.03 & 1.4$\pm$0.2 & 8.6 & 653$\pm$30 & 382.2$\pm$47.4 & 110$\pm$12 & 0.4$\pm$0.3 & 870$\pm$600 & 4.4$\pm$0.6  & 40$\pm$30& 0.7$\pm$0.4 \\
3C 318 & 0.25$\pm$0.03 & 3$\pm$0.3 & 20.2 & 1132$\pm$44 & 528.7$\pm$67.4 & 168$\pm$18 & 0.32$\pm$0.2 & 220$\pm$150 & 12$\pm$2 & 10$\pm$7  & 8$\pm$3 \\
3C 298 & 0.3$\pm$0.03 & 3$\pm$0.3 & 1.6 & 394$\pm$64 & 624.0$\pm$49.0 & 780$\pm$150 & 0.6$\pm$0.3 & 750$\pm$400 & 20$\pm$7 & 77$\pm$40  & 4$\pm$2 \\
4C 09.17 A & 0.11$\pm$0.01 & 1.3$\pm$0.1 & 2.8 &852$\pm$77& 439.1$\pm$122.6 & 400$\pm$50 & 0.05$\pm$0.02 & 50$\pm$20 & 21$\pm$4 & 2.1$\pm$1  & 2.4$\pm$0.5 \\
4C 09.17 B & 2.3$\pm$0.2 & 27$\pm$3 & 4.9 & 456$\pm$26  & 870.6$\pm$47.2 & 2500$\pm$300 & -- & -- & 73$\pm$11 & --  & -- \\
\enddata
\end{deluxetable*}

\section{Individual Objects}\label{sec:indiv_obj}

In this section, we outline the known properties of each quasar within our sample, focusing on their radio jet morphology, far-infrared properties, morphologies, the extent of the ionized gas outflows, and a description of what we detect with our ALMA band 4 observations.

\subsection{3C 318 (z=1.5734)}\label{sec:3c318}

3C 318 is a luminous radio-loud quasar at $z=1.5734$. The radio jet is double-sided, stretching in the southwest and northeast, with a bright core emission associated with the quasar's optical emission location. Within our ALMA continuum observations, we only detect the jet's southwest component. The northeast component of the jet blends with the bright unresolved core. 3C 318 has been detected with the \textit{Herschel Space Telescope} and is known to be a bright far-infrared emitting source \citep{Podigachoski15}. Resolved band 7 ALMA observations of the dust emission reveal a ring-like structure on kpc scale centered on the quasar \citep{Barthel19}.\\

With the Keck/OSIRIS, we detected ionized gas emission in nebular emission lines \oiii 5007 \AA, \ha, \nii 6585 \AA, and \sii 6717, 6731 \AA\ with an extent of 4 kpc \citep{Vayner19b}. We detected an ionized outflow extending in the SW and NE direction with a maximum extent of 3.2 kpc. \\

3C 318 was known to have molecular emission detected at the CO (2-1) transition with PdBI \citep{willott07}. The CO (2-1) emission was known to be blueshifted by 400 \kms\ and spatially offset from the quasar continuum by 2.4\arcsec\ to the west and 0.5\arcsec\ to the north with considerable uncertainty due to a coarse beam of 8.05\arcsec$\times$4.32\arcsec. \\

\begin{figure*}[!th]
    \centering
    \includegraphics[width=7.0in]{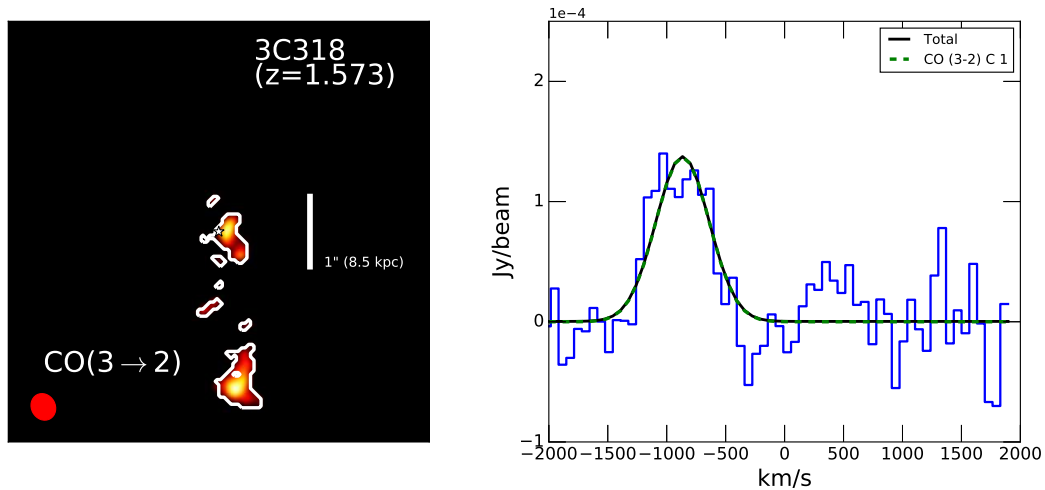}
    \caption{ALMA band 4 observations of 3C 318. On the left we show optimally extracted intensity map of the molecular outflow in the 3C 318 system, detected in the CO (3-2) line. The white contours outline the molecular outflow region. On the right we show a spectrum integrated over the entire molecular outflow region shown in the white contours, along with fit to the CO (3-2) emission line. The systemic redshift of the quasar host galaxy is at 0 \kms. The ellipse in the lower left corner on the right panel shows the beam of our ALMA band 4 observations. We detect no molecular CO (3-2) emission at the systemic redshift.}
    \label{fig:3C318_spec}
\end{figure*}

Our ALMA band 4 observations reveal an extended CO (3-2) emission with one component offset 1.7 kpc to the west and a second component offset 17 kpc towards the south. The emission is divided between two regions that have widths of 4 and 8 kpc, that show similar highly blueshifted emission relative to the quasar. The spatially integrated emission is blueshifted (-936.0 \kms) and relatively broad with an FWHM of 534 \kms. The spectrum and integrated intensity map of the detected CO emission are shown in Figure \ref{fig:3C318_spec}. Likely the ALMA and PdBI observations trace the same molecular gas components. However, the differences in the beams, maximum recoverable scales, and SNR play a role in the observed line shift and integrated line flux. 3C 318 has also been observed with VLA targeting the CO (1-0) transition \citep{Heywood13}. Emission associated with the CO (1-0) emission line at the velocity offset of the PdBI observations was found 0.33\arcsec\ north from the quasar continuum. We do not detect any CO (3-2) emission at that location. Using typical ratios between the CO (1-0) and CO (3-2) line luminosity, we would have expected to detect this component at an SNR of 100, integrated over an emission line equivalent with a velocity dispersion of 250 \kms. \\

The separation between the two high velocity clumps roughly matches the maximum recoverable scale of the interferometric observations. We have attempted to recover the more diffuse emission between the two clumps by smoothing the data in the UV plane with Gaussian kernel using the \textit{uvtaper} option within \textit{tclean} in CASA. Using a uv-taper parameter of 0.5\arcsec\ and 1\arcsec\ on the sky, the fainter emission between the two clumps seen in Figure \ref{fig:3C318_spec} is below the noise in the uv-tapered data. The total integrated flux from the CO (3-2) emission is within 10\% of the original data, within the statistical noise of the observations, hence no additional ``diffuse" emission was recovered. A loss of baselines resulted in an increase in noise with larger uv-taper parameters. Observations in a more compact ALMA configuration are necessary to detect the more diffuse emission. \\

ALMA's higher angular resolution and sensitivity lead us to speculate that the molecular gas emission in 3C 318 is associated with a molecular outflow rather than a merging system. This is supported by more accurate redshift measurement from the narrow-line region, allowing us to do better kinematics and dynamics measurement of the molecular gas emission. The Dark Energy Survey \citep{DES2021} shows no apparent optical detection of a companion galaxy down to an $r$-band magnitude of 24 at a significance of 10$\sigma$. Infrared observations with Spitzer and archival ALMA band 7 observations do not show any evidence for a galaxy at the spatial locations of the highly blueshifted CO (3-2) emission \citep{Barthel19}. There may be a possibility that high-velocity emission is associated with an obscured galaxy near the 3C 318 quasar host galaxy. In recent years there have been several galaxies detected with ALMA that have faint or no counterparts in very deep optical imaging and are referred to as ``optically dark ALMA galaxies" \citep{Williams19,Zhou20}. These galaxies show high far-infrared luminosities and are characterized as dusty star-forming galaxies at high redshift ($z>2$). Such galaxies are typically relatively massive, with stellar masses of 10$^{10.2-11.5}$ \msun\ and contain a substantial amount of molecular gas. If the emission in 3C 318 is associated with such a galaxy, then the associated dust continuum emission would have been easily detectable in the ALMA band 7 observations.\\ 

We divide the molecular gas mass by the area of the emitting region using the effective radius as the scale size. Using the Milky Way's hydrogen column density - $V$-band extinction relationship \citep{Guver09} we convert the column density into a $V$-band extinction value for the molecular gas traced by CO. We measure an $\rm A_{v}$ value in the CO gas of 1-4 magnitudes, which is a factor of 100 lower than those found in the dusty star-forming galaxies. Furthermore, we use the molecular gas mass of individual clumps and convert them into expected dust continuum emission in band 7 based on the relationship between ISM mass and dust emission of \citet{Scoville17}. The expected continuum flux density is 0.017 mJy/beam, which would be undetected in the band 7 observations with a sensitivity of 0.0297 mJy/beam. We assume the clumps have the same surface brightness profile in both bands and in the dust and molecular gas emissions. Based on these calculations, the clumps are unlikely to be associated with typical dust-obscured galaxies at this redshift. Band 4 continuum is not optimal for detecting the dust continuum at this redshift since the dust emission is expected to be fainter at the longer wavelength of the Rayleigh-Jeans tail. Another possibility is that the emission may be associated with a tidal tail feature. However the FWHM of 534 \kms\ and an offset of -936.0 \kms\ relative to the redshift of the quasar are both larger than what would be expected for a tidal feature. Based on morphological identification of tidal tail from HST imaging and Keck/OSIRIS observations, in two other systems, we found that the velocity dispersion in both the ionized and molecular gas mass is $\sim$ 150 \kms\ with velocity offsets of -250 \kms\ \citep{Vayner19b}. \\

Combining rest-frame optical and sub-mm observations, we find that the ionized and molecular gas outflows in 3C 318 show different morphologies, spatial extent, and kinematics. The molecular gas outflow is far more extended, with a maximum distance of 21 kpc from the quasar, while the ionized outflow shows a maximum extent of only 3.2 kpc. The molecular outflow is also faster moving with blueshifted velocities up to -1200 \kms. In contrast, the ionized gas outflow has a velocity of 703 \kms\ with a bi-conal morphology that is both blue- and red-shifted relative to the quasar in the SW and NE directions. We find no evidence for CO (3-2) emission at the quasar's systemic redshift associated with narrow emission. We place a limit on the molecular gas reservoir at the quasar's location over an aperture matching the size of the beam of $<0.7\times10^{9}$ \msun\ at 2$\sigma$ confidence.\\

The extent of the molecular outflow is similar to the cold gas outflow detected in $z=6.4$ quasar SDSS J1148+5251, through the 158 \micron\ [C II] emission line \citep{Cicone15}. The morphology and kinematics of the molecular outflow in 3C 318 are also similar to the outflow recently detected in zC400528 \citep{Herrera-Camus19} through CO (3-2) observations, where they see an extended emission entirely redshifted from the galaxy with a relatively collimated morphology similar to the case of 3C 318. The clumpy morphology and high velocity of the outflowing molecular gas are also similar to recent molecular outflows detected in PDS 456 \citep{Bischetti19} and in the lensed quasar HS 0810+2554 \citep{Chartas20}.

\subsection{4C 09.17 (z=2.117)}\label{sec:4c0917}

4C 09.17 is a radio-loud quasar at $z=2.117$. A one-sided jet extends towards the southwest, with a bright core emission associated with the quasar's optical emission location. The system is also bright at far-infrared wavelength \citep{Podigachoski15}. \\

With Keck/OSIRIS we detected ionized gas emission in nebular emission lines \oiii\ 5007 \AA, \ha, and \nii\ 6585 \AA\ \citep{Vayner19b}. We found an ionized gas outflow extending towards the east with a maximum extent of 6 kpc.\\ 

We detect broad, blueshifted emission resembling a molecular gas outflow in the host galaxy of the radio-loud (RL) quasar 4C 09.17. From here on, we refer to this object as 4C 09.17 A - RL. We also detect a very broad component in the merging radio-quiet (RQ) galaxy towards the northeast; from here on, we refer to this object as 4C 09.17 B - RQ. This galaxy is also detected in \textit{K}-band imaging of \citet{Armus97}, with a red optical to near-IR continuum color. In 4C 09.17 B - RQ, we detect very faint narrow \oiii\ emission in Keck/OSIRIS, the ionized outflow is undetected. 4C 09.17 B - RQ also contains a narrower emission line component in CO (4-3) at a similar redshift to the narrow \oiii\ emission, which we use to calculate its redshift. The velocity offset between the 4C 09.17 A - RL and 4C 09.17 B - RQ is -593 \kms. The majority of the narrow CO emission-line flux is found concentrated in 4C 09.17 B - RQ, within a 1 kpc radius region. The majority of the dust continuum detected at far-infrared wavelengths with the \textit{Herschel Space Telescope} is likely associated with this galaxy. 4C 09.17B - RQ is highly obscured; the narrow CO emission component yields a line integrated gas column density of 3.4$\rm \times10^{24}~cm^{-2}$ computed by dividing the molecular gas mass by area of the emitting region. Using the Milky Way's hydrogen column density - $V$-band extinction relationship \citep{Guver09}, we find a $V$-band extinction of 150 mag. The narrow CO emission is likely at the center of the merging galaxy because it roughly corresponds to the $K$-band continuum's peak location.\\

For the galaxies detected to the southwest and northwest of the quasar in \citet{Armus97} and \citet{Lehnert99} we detect narrow CO (4-3) emission near their optical locations. We do not detect any high velocity or broad molecular gas associated with these two systems. The detection of 3 galaxies found within 20 kpc of the quasar host galaxy from both ALMA and Keck/OSIRIS observations makes the 4C 09.17 system likely a proto-group environment at z $\sim 2.11$. \\

We find that the molecular gas outflow in 4C 09.17 A -RL is more compact than the ionized gas outflow. The ionized and molecular gas outflow show similar blueshifted velocities and velocity dispersion. The maximum extent of the molecular gas outflow is 2.8 kpc, while the ionized outflow extends to 6 kpc. We find that both the ionized and molecular outflow in this system are not along the path of the radio jet, but extend in the same eastern direction. Similar results have been found for a subset of nearby galaxies recently, where outflows appear to expand perpendicular to the path of the jet \citep{Venturi21}. \\

\begin{figure*}
    \centering
    \includegraphics[width=6.5 in]{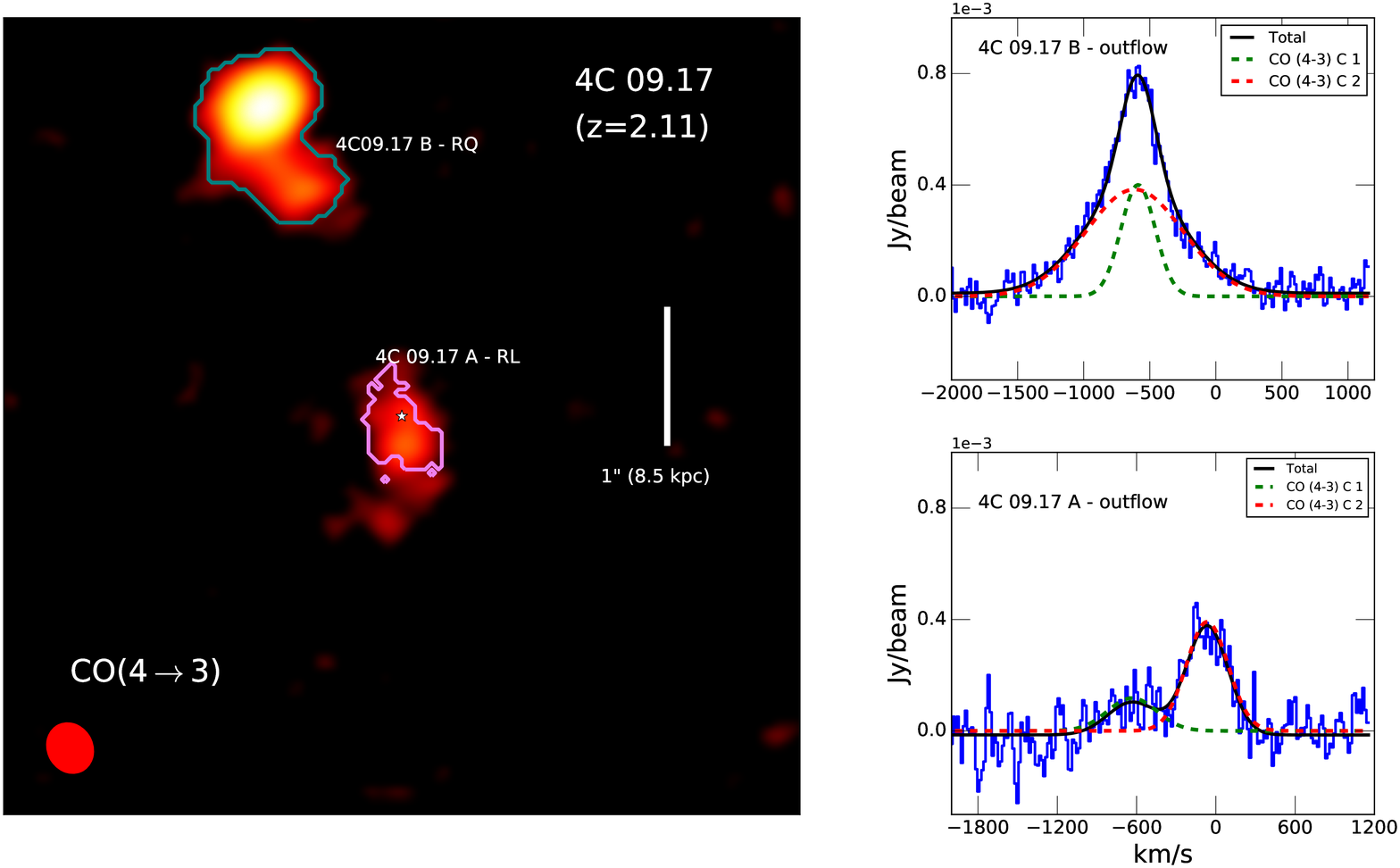}
    \caption{ALMA band 4 observations of 4C 09.17. On the left we show optimally extracted intensity map of CO emission in the 4C 09.17 system. The 4C 09.17 is a merger system with molecular outflows detected in both galaxies. The teal contours outline the molecular outflow in the radio-quiet galaxies 4C 09.17 B, while the purple contours outline the molecular outflow detected in the host galaxy of the radio loud quasar 4C 09.17. On the right we show the spectra extracted over the respective outflow regions along with the Gaussian fit model. Dashed lines represent the individual Gaussian components of the emission line fit, while the solid black line represents the sum of all components and a 0th order polynomial fit to any residual continuum. Component 1 (C1) in 4C 09.17 B is the fit to the narrow emission at the systemic redshift of the merging galaxy, that has a velocity offset of about 593 \kms\ relative to 4C 09.17A, while C2 is gas in the outflow. For 4C 09.17 A, C1 corresponds to the outflow gas while C2 is the narrow gas at the systemic redshift of the quasar host galaxy. The systemic redshift of the quasar host galaxy is at 0 \kms. The ellipse in the lower left corner on the right panel shows the beam of our ALMA band 4 observations.}
    \label{fig:4C0917_spec}
\end{figure*}

In 4C 09.17 B - RQ, the molecular outflow extends 4.9 kpc from the narrow CO emission line component. The extent of the molecular outflow in 4C 09.17 B - RQ roughly matches the maximum extent of the $K$-band stellar continuum; hence the molecular outflow is occurring on galactic scales in this galaxy. Extinction within the outflowing gas can potentially prevent ionization by quasar photons and prevent the observer from detecting recombination photons. Spectra of the distinct regions detected in this system are shown in Figure \ref{fig:4C0917_spec}.  \\

\subsection{4C 22.44 (z=1.5492)}\label{sec:4c2244}

4C 22.44 is a radio-loud quasar at z=1.5492. The radio jet in this system extends along the east-west direction with a length of 7\arcsec\ and a bright core emission associated with the quasar's optical emission location. With Keck/OSIRIS we detected extended ionized gas in the \oiii, \ha, and the \nii emission lines with an extent of 2 kpc. An ionized gas outflow is detected on a spatial scale of $<1$ kpc. We detect no emission in the CO (3-2) line with ALMA. We place a limit on the CO (3-2) line luminosity of 0.08 Jy \kms\ for an aperture matching the beam with a line velocity dispersion of 250 \kms, which converts to a molecular gas mass limit of $<$ 1$\times10^{9}$ \msun.

\subsection{4C 05.84 (z=2.323)}\label{sec:4c0584}

4C 05.84 is a radio-loud quasar at z=2.323. The one-sided radio jet in this system extends towards the southwest with a maximum extent of 12 kpc. There is a bright radio core component associated with the quasar's optical emission location.

With Keck/OSIRIS observations, we detected extended ionized gas on an 8 kpc scale in the \oiii, \ha, and \nii emission lines \citep{Vayner19a}. We detected a bi-conical ionized outflow extending along the northeast and southwest direction.

\begin{figure*}[!th]
    \centering
    \includegraphics[width=7.5 in]{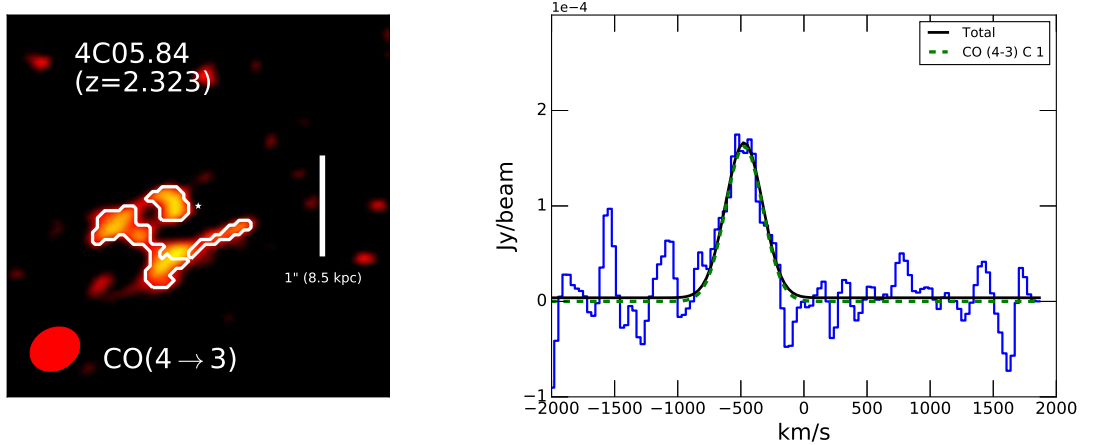}
    \caption{ALMA band 4 observations of 4C 05.84. On the left we show optimally extracted intensity map of the molecular outflow in the 4C 05.84 system, detected in the CO (4-3) line. The white contours outline the molecular outflow region. On the right we show a spectrum integrated over the entire molecular outflow region, along with fit to the CO (4-3) emission line. The systemic redshift of the quasar host galaxy is at 0 \kms. The ellipse in the lower left corner on the right panel shows the beam of our ALMA band 4 observations. We detect no molecular CO (4-3) emission at the systemic redshift.}
    \label{fig:4C0584_spec}
\end{figure*}

In our ALMA band 4 observations, we detect extended CO (4-3) emission that is offset in the western direction consisting of multiple clumps. The individual clump components and the spatially integrated emission is highly blueshifted relative to the quasar's systemic redshift. This emission is not associated with any known galactic component. The spectrum and integrated intensity map of the detected CO emission are shown in Figure \ref{fig:4C0584_spec}. 

The ionized and molecular gas outflows in this system are on similar scales. The ionized outflow extending towards the southwest direction shows a similar blueshifted velocity offset as the molecular outflow. The ionized outflow detected on spatial a scale $<1$ kpc is similarly blueshifted to the more extended ionized and molecular outflow. The ionized outflow appears to be more turbulent with a larger velocity dispersion than the molecular outflow. We find no evidence for CO (4-3) emission at the quasar's systemic redshift associated with narrow emission. We place a limit on the molecular gas reservoir mass of $<0.8\times10^{9}$ \msun\ at the quasar's location with an aperture the size of the beam and using a velocity dispersion of 250 \kms.

\subsection{3C 298 (z=1.439)}\label{sec:3C298}

3C 298 is a radio-loud quasar at z=1.439. The jets in the system extend in the east-west direction, with a length of about 16 kpc and a bright core emission associated with the quasar's optical emission location. The system is bright at far-infrared wavelengths. ALMA band 7 observations have revealed that the dust emission mostly comes from a kpc scale region centered on the quasar \citep{Barthel18}.

With Keck/OSIRIS, we detected extended ionized gas emission \citep{Vayner19b} on scales up to 20 kpc from the quasar. A bi-conical ionized outflow is detected with a maximum extent of 3 kpc from the quasar along the jet's path. The redshifted cone is associated with the western jet component, while the blueshifted cone is associated with the eastern jet component. We also detected an ionized gas outflow in the merging system 8 kpc from the quasar.

In the ALMA band 4 observations, we detect a molecular gas emission in two distinct components, one centered on the quasar with a radius of 2.1 kpc and a second component 21 kpc from the quasar, offset by -250 \kms. The component near the quasar shows both broad and narrow emission. The narrow emission is associated with a galactic disk \citep{Vayner17}, while the broad component is associated with the outflow. The outflow emanates from the molecular disk centered on the quasar with a maximum extent of 1.6 kpc. The majority of the molecular gas in the outflow is on the blueshifted side of the disk and extends in the direction of the jet's western component. The ionized outflow is more extended than the molecular outflow, has a faster velocity, and appears to be more turbulent with a larger velocity dispersion.

\subsection{7C 1354+2552 (z=2.0064)}\label{sec:7C1354}

7C 1354+2552 is a radio-loud quasar at z=2.0064. The system contains two jets that are perpendicular to each other. The east-west jet has a length of about 24 kpc, while the north-south jet has a length of 86 kpc. Only the east-west jet is detected in the continuum in our ALMA observations due to limited sensitivity to low-surface brightness emission on scales $>$ 6\arcsec.

Using the Keck/OSIRIS observations, we detected extended emission in the nebular emission lines \oiii and \ha on scales of 4-6 kpc. An ionized outflow is detected on a spatial scale of $<1$ kpc. 

\begin{figure*}[!th]
    \centering
    \includegraphics[width=7.0 in]{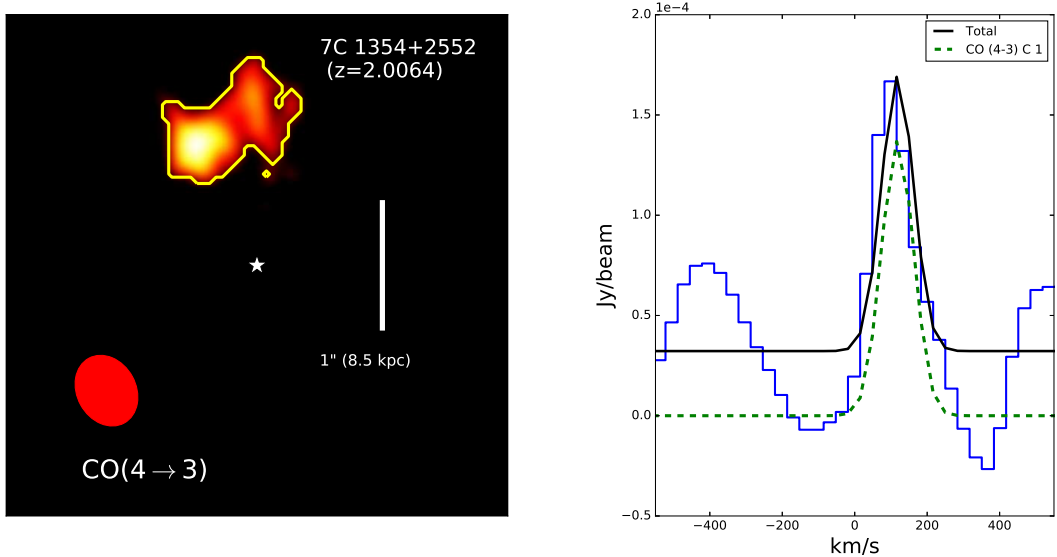}
    \caption{ALMA band 4 observations of 7C 1354+2552. On the left we show optimally extracted intensity map of the molecular gas in the 7C 1354+2552 system, detected in the CO (4-3) line. The gas is narrow and is consistent with gravitational motion. Yellow contours outline the CO emitting region towards the north-west that is slightly redshifted, the spectrum associated with it is shown on right. The systemic redshift of the quasar host galaxy is at 0 \kms. The ellipse in the lower left corner on the right panel represents the beam of the ALMA band 4 observations.}
    \label{fig:7C1354_spec}
\end{figure*}

In the ALMA band 4 observations, we detect molecular gas emission towards the northeast. The narrow emission line resides near the quasar's systemic redshift. We do not detect any broad emission from a turbulent molecular gas outflow. We detect no highly offset emission consistent with our outflow criteria; hence there is no evidence for a cold molecular gas outflow in this system at our observations' sensitivity. Spectra of the distinct region detected in this system is shown in Figure \ref{fig:7C1354_spec}. The detected emission may be associated with a merging galaxy. With the large (8 kpc) separation from the quasar and the fact that the motion of the gas does not appear to follow the kinematics of the galactic disk on the eastern side \citep{Vayner19b}, there is a possibility that this emission is associated with a merging galaxy.

\section{Discussion}\label{sec:discussion}

\subsection{What is driving the outflows?}\label{sec:driving}
Several powerful mechanisms can drive galactic-scale outflows. Quasars can reside in galaxies with powerful star formation activity \citep{Duras17,Aird19,Circosta21,Bischetti21}, especially at redshifts near the peak of the star formation activity. Star formation can result in powerful galactic winds \citep{Rupke18}. Radio-loud quasars that are optically luminous can drive galactic outflows by both jets \citep{Wagner12,Mukherjee16} and radiation pressure \citep{Murray95,Faucher12a,Zubovas12,Costa18} from the accretion disk. In this section, we explore primary mechanisms that are capable of driving the multi-phase galactic outflows. We combine the momentum flux and kinetic luminosities from the cold molecular and ionized gas phases to look at both the total impact of galactic winds on the quasar host galaxies and what mechanism may be responsible for driving the entirety of the outflowing gas in each system.

To understand the main driving mechanism of galactic outflows, we compare the momentum flux of the outflow (\momfluxout) to the momentum input from the quasar accretion disk (\momfluxagn) and to the momentum deposition from stellar feedback ($\dot{P}_{SNe}$).

\subsection{Stellar feedback as potential driver of galactic outflows}
To explore whether star formation activity can drive the galactic scale outflows, we compare the momentum deposition from supernovae explosions, based on the star formation rate, to the energetics of the multi-phase gas outflow. To compute the energy deposition from supernovae, we use the results of recent simulations by \citet{2015MNRAS.450..504M} and \citet{2015ApJ...802...99K}, that predict the momentum deposition per unit of solar mass formed. We assume that one supernova explodes for every 100 \msun\ of stars. We assume an electron density of 100 \eden in the interstellar medium with solar metallicity.

\begin{equation}
    \dot{P}_{SNe} = 7.5\times 10^{33} \frac{\dot{M}_{SFR}}{1M_{\odot}{\rm yr}^{-1}} \left(\frac{n}{100\,{\rm cm}^{3}}\right)^{-0.18}
    \left(\frac{Z}{Z_\odot}\right)
    \rm dyne \label{eq:SNe_mom}.
\end{equation} 

For sources with resolved rest-frame far-infrared observations (3C 318 and 3C 298), we use the far-infrared luminosity and convert that into a star formation rate based on work by \citet{Barthel18,Barthel19}. The total-infrared (8-1000 \micron) derived star formation rates of 3C 298 and 3C 318 are 930\myr\ and 580 \myr\ derived using the \citet{Kennicutt98} calibration \citep{Podigachoski15}. Infrared-derived star formation rates have high uncertainties since it is not clear what fraction of the far-infrared emission comes from dust heating from massive stars and quasar heating on kpc scales \citep{Symeonidis17,Symeonidis21}. In radio-loud objects synchrotron emission from the jets and the core of the quasar can also contribute to the far-infrared and mm-emission. In \citet{Podigachoski15} correction for synchrotron emission was only applied to the 850 \micron\ flux of a handful of objects, including 3C 298 that is part of our study. We therefore use the infrared-derived star-formation rates as an upper limit on the actual star formation rates. The total infrared star formation rates are presented in Table \ref{tab:SFR-rates}. \\

Another tracer of recent star formation comes from recombination lines of hydrogen. In Table \ref{tab:SFR-rates} we present the star formation rates derived from the \ha line luminosity using the \citep{Kennicutt98} calibration. The \ha derived star formation rates are derived from integrated \ha emission at the systemic redshift of the quasar host galaxy. The \ha and total infrared derived star formation rates show a significant discrepancy, with the far-infrared star formation rates being almost an order of magnitude higher. The discrepancy between these inferred star formation rates may be due to several reasons; dust extinction, different tracers of the star formation history, contamination of the far-infrared emission from quasar processes, or in addition a large difference between the resolution of the \textit{Herschel Space Telescope} and Keck/OSIRIS observations. Indeed, the far-infrared derived star formation rate in 4C 09.17 can be attributed to several galaxies falling within \textit{Herschel Space Telescope} PSF leading to contamination of the far-infrared emission. The merging galaxy 4C 09.17 B-RQ contains higher molecular gas surface density, hence the larger fraction of the star formation rate among the galaxies. Contribution from nearby galaxies can also affect the far-infrared derived flux for 3C 298 and 3C 318, but we do not see such a strong over-density compared to 4C 09.17. Recent observations with ALMA of a high redshift ($z=4.4$) and luminous quasar have revealed multiple sources falling within 17 kpc from the quasar \citep{Bischetti18}, which all contribute to the far-infrared emission detected with \textit{Herschel Space Telescope} for this system. 

We find that stellar processes can deposit a momentum flux of 7.5 - 10,000 $\times10^{33}$ dyne, comparable to momentum fluxes of the multi-phase outflows. However, since these estimates rely on the maximum momentum flux from SNe and the maximum star formation rates, it is unlikely that star formation alone drives the outflows in these systems.

\begin{deluxetable*}{cccc}
\tiny
\tablecaption{Star formation rates based on \ha emission line \citep{Vayner19b},infrared observations \citep{Podigachoski15} and expected star formation rate using the molecular gas surface density based on the KS law. \label{tab:SFR-rates}}
\tablehead{
\colhead{Name} & 
\colhead{SFR [\ha]}&
\colhead{SFR [Total IR]}&
\colhead{SFR [KS]}\\
\colhead{} & 
\colhead{\myr}&
\colhead{\myr}&
\colhead{\myr}}
\startdata
4C 09.17 A RL &  9$\pm$1 & 1330 \tablenotemark{a}  & 6  \\
7C 1354+2552 & 	 29$\pm$3 & -- & 1 \\
3C 298 &  22$\pm$2 & 930 & 24 \\
3C 318 &  88$\pm$9 & 580 & - \\
4C 22.44 &  32$\pm$3 & -- & - \\
4C 05.84 &  11$\pm$1 & $<$540 & - \\
\enddata
\tablenotetext{a}{Star formation rate likely contaminated by a merging galaxy}
\end{deluxetable*}

\subsection{Quasar as potential driver of galactic outflows}
The photon momentum fluxes of $10^{35-37}$ dyne and bolometric luminosities of $10^{46-47}$ \ergs\ for these quasars indicate that they have sufficient energy and momentum to drive the observed multi-phase gas outflows. To test which quasar mechanism is the primary driver of the observed galactic-scale outflows, we compare \momfluxout to \momfluxagn and the location and extent of the quasar jets. High ($>2$) \momfluxratio\ on scales $>$ 1 kpc are generally attributed to energy-conserving outflows, where a radiatively-driven nuclear wind \citep{Faucher12a} or a jet \citep{Mukherjee16,Wagner12} drives a hot shock (T$>10^{7}$ K) in the interstellar medium that does not cool efficiently, maintaining nearly all of the kinetic energy provided to it. The swept-up material is shocked and is able to cool to produce a multi-phase outflow medium \citep{Faucher12,Faucher12a,Zubovas12}, and may explain the presence of molecular gas moving at fast outflow velocities. To decipher whether the galactic-scale wind is ultimately powered by a jet or by radiation, we need to compare the location and extent of the quasar jet to the outflow and to search for evidence of a fast, radiatively driven wind in X-ray and UV spectrum of the quasar.

High ($>2$) \momfluxratio on scales $<$ 1 kpc can be attributed to outflows driven by radiation pressure in a high column density environment, where the ``momentum boost" is provided by photons scattering multiple times off dust grains as they are driving the outflow \citep{Thompson15}. Detecting and resolving the hot shocked gas produced by the jet or the radiatively-driven nuclear wind would be helpful in understanding what is driving the outflow. The shocked hot gas can be detected through the Sunyaev–Zeldovich effect \citep{Chatterjee07} and has recently been found in the host galaxy of a luminous quasar at z=1.71 \citep{Lacy19}. Additionally it was detected by stacking analysis of luminous quasars from the Atacama Cosmology Telescope, \textit{Herschel Space Telescope} and VLA observations \citep{Hall19}.

Low ($\ll1$) \momfluxratio\ on scales $>$ 1 kpc can be attributed to outflows driven by a radiative shock produced where the shocked gas can cool efficiently, and kinetic energy is radiated away. Outflows driven through radiation pressure in a low column density environment can also produce an observed \momfluxratio $\ll$1. 

\begin{figure*}
    \centering
    \includegraphics[width=7.0 in]{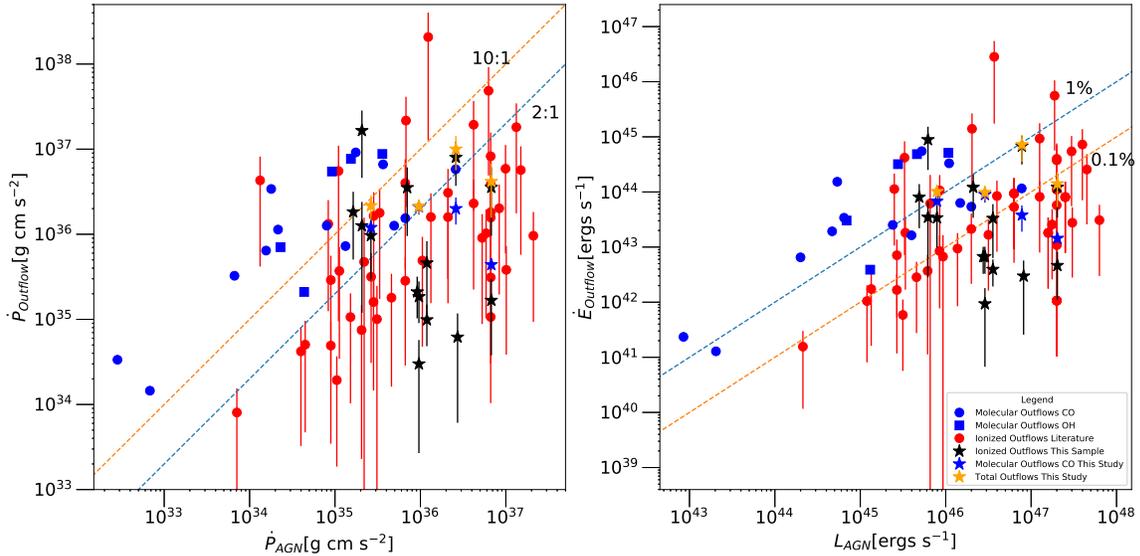}
    \caption{A comparison of the momentum flux (left) and kinetic luminosity (right) of the outflow to the momentum flux of the accretion disk and the quasar's bolometric luminosity. On the left, we plot lines of constant \momfluxratio, points above the 2:1 line represent outflows that are likely driven by an energy-conserving shock on kpc scale or radiation pressure on small ($<1$ kpc) scales. On the right, we plot lines of constant coupling efficiency between the outflow's kinetic luminosity and the quasar's bolometric luminosity. Blue circles and squares represent molecular outflows detected through CO emission and OH absorption, respectively. Red circles represent ionized outflows at cosmic noon, recomputed in the exact same manner in \citep{Vayner19a}. Black stars are the ionized outflows of the parent sample for this study. Blue stars are molecular outflows derived in this study through CO emission, and orange stars are the energetics from the total (molecular + ionized) momentum flux and kinetic luminosity. Combining the energetics of the multi-phase outflows indicates that they are likely driven by an energy-conserving shock and have coupling efficiencies between 0.1-1\%.}
    \label{fig:energetics}
\end{figure*}

\begin{figure*}
    \centering
    \includegraphics[width=7.0in]{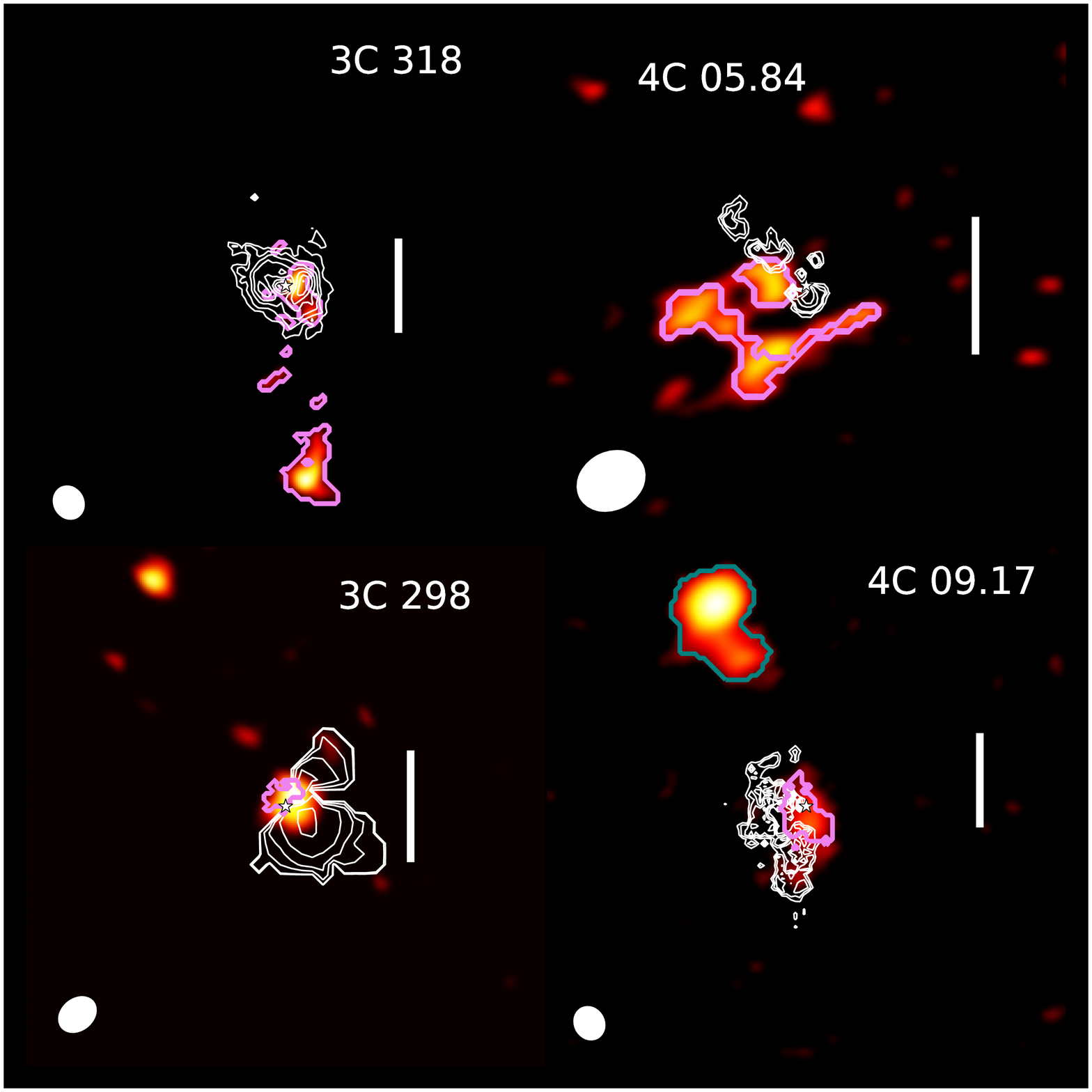}
    \caption{We present the comparison between the ionized outflow and the molecular gas distribution, extent, and morphology in sources where we detect a multi-phase gas outflow. In the background, we show the line integrated CO intensity tracing the cold molecular gas. Violet contours outline the location of the molecular outflows in the quasar host galaxies, the teal contour in the 4C09.17 system represent the molecular outflow in the merging galaxy. The white contours show the ionized outflow traced through the \oiii\ 5007 \AA\ emission line for all systems except 4C 05.84 where it is traced with \ha. The white star represents the quasar's location, while the bar to the right of each source represents 1 arcsecond.}
    \label{fig:CO_ionized_outflows}
\end{figure*}

To compare the energetics of winds of our sample with other massive galaxies with AGN, we collated data on molecular and ionized outflows in galaxies at $1<z<3$. We recomputed the ionized outflow rates, luminosities, and momentum fluxes using a consistent prescription; details can be found in \citet{Vayner19a}. To differentiate AGN outflow mechanisms we compare the momentum flux of the outflow against the photon momentum flux from the accretion disk, as well as and the kinetic luminosity of the outflow against the bolometric luminosity of the quasar (Figure \ref{fig:energetics}). On Figure \ref{fig:energetics} we overlay theoretical predictions of the expected \momfluxratio value based on energy-conserving outflows, radiative shock driving, and radiation pressure \citep{Faucher12,Zubovas12,Zubovas14}. Minimum coupling efficiency between the quasar bolometric luminosity and outflow kinetic luminosity are prescribed in theoretical simulations when outflows directly impact the star-forming properties of the host galaxy. Based on theoretical models, minimum coupling efficiencies are presented in Figure \ref{fig:energetics} \citep{Hopkins10,Choi12,Costa18}. We present the molecular outflows and the total energetics from the multi-phase outflow in Figure \ref{fig:energetics}. After combining the ionized and molecular outflows' energetics, the outflows exhibit a \momfluxratio $>1$ on kpc scales. An energy-conserving shock is responsible for driving the outflow for 3C 318, 4C 09.17 A, and 3C 298. For 4C 05.84, it is still possible for an energy-conserving shock to drive the outflow; however, radiation pressure or a radiative shock model is still able to explain the observed \momfluxratio. We overlay the location of the ionized and molecular outflows in Figure \ref{fig:CO_ionized_outflows}, to highlight the extent and morphologies of the multi-phase gas outflows.

\subsection{The nature of the multi-phase outflows}

For 3C 318, 4C 05.84, and 4C 09.17 A, compared to the ionized outflowing gas the molecular gas shows a smaller velocity range single component that spans only blueshifted emission from -200 to -1200 \kms. The morphology of the outflowing molecular gas is much clumpier and more confined compared to the ionized outflows. The ionized outflows span a broader range of velocities and fill a larger volume likely residing in a wide-angle cone. The high velocity and clumpy molecular outflows are unexpected, appearing to be out of pressure equilibrium with their surroundings at the larger separations observed in 3C 318 and 4C 05.84. Recently high velocity and clumpy outflows have also been detected in one low redshift radio-quiet quasar and a higher redshift lensed quasar, appearing to have similar velocity offset, dispersion, and morphology to 3C 318, 4C 05.84 and 4C09.17A \citep{Bischetti19,Chartas20}. There is no consensus on the observed spatial morphology and velocity structure of high redshift molecular outflows in luminous quasars due to the small number of detections. The velocity dispersions and offsets for 3C 318, 4C 05.84, and 4C 09.17 A are consistent with molecular outflows detected through OH absorption \citep{Veilleux13}, which have been hypothesized to have a thin shell geometry. For 3C 298 and 4C 09.17 B, the molecular outflows show a broader velocity range, which likely indicates that these molecular outflows are more volume filling, similar to the ionized outflows. In case these outflows are indeed filled cones, then their outflow rates and energetics would scale by a factor of 3. We are unable to fully rule out a shell geometry for these outflows. It would require higher angular resolution observations to distinguish if they reside in a shell or in a filled cone/sphere. The velocities of the molecular and ionized outflows are consistent with recent theoretical predictions by \citet{Richings20} for multi-phase gas outflows driven by a quasar. \\

The energy-conserving shocks responsible for the galactic scale outflows are either powered by radiatively driven outflow or through shock heating of the gas by the radio jet. In 3C 298 and 3C 318, the jet's path is consistent with both the ionized and molecular outflows, while in 4C 05.84, the jet is only consistent with the path of the ionized outflow; however, the size of the jet is still consistent with both the extent of the ionized and molecular gas outflows. In 4C 09.17, the jet is not consistent with either the molecular or ionized outflow path. The lack of correlation between the jet and the galactic outflows does not indicate that the jet could not have driven the outflow. At later stages in their evolution, the hot bubble cocoons shocked by the jet could be relatively spherical and volume filling relative to the thin jet that we observe at radio wavelength \citep{Wagner12,Mukherjee16}. The lack of path correlation between the multiple phases of the outflow, the general asymmetry, and spatial differences in ionized and molecular outflow can be caused by the non-uniform cooling properties of the galactic outflow, subject to gas and dust extinction and surface brightness sensitivity of our observations. Some nearby galaxies show fast-moving outflows that are not correlated with the axis of the jet \citep{Venturi21} but are rather found to expand perpendicular to the jet's path. None of the quasars appear to show evidence for a broad absorption line wind, based on their rest-frame UV spectra from SDSS \citep{Paris18}. The X-ray observations are either missing or are too shallow to search for radiatively-driven nuclear winds in Fe absorption lines. Future space-based X-ray telescopes with larger effective apertures and higher spectral resolving power will be crucial to search for ultra-fast outflows and compare their energetics with the galactic scale winds to decipher whether the jet or quasar disk winds are the primary driver.

At the current evolutionary stage of the quasar host galaxies in our study, the dominant outflow component, in terms of mass, is the molecular gas. In fact, for sources with both a molecular outflow and a molecular gas reservoir at the systemic redshift, between 20-60\% of the cold molecular gas reservoir is in the galactic outflow. For sources with no detected molecular reservoir at the systemic redshift, we are catching these systems when the majority of the cold molecular gas is in an outflow. This suggests that we are observing these systems at a phase where the quasar is responsible for removing a significant fraction of the gas in these galaxies, and therefore may be directly responsible for removing the fuel for subsequent star formation, in turn impacting the stellar growth of these galaxies.

For the galactic outflows in 3C 298 and 4C 05.84, we find that most of the momentum flux is in the ionized gas phase. For 3C 318, the ionized outflow can have higher energetics within the errors that are dominated by the measurement of the electron density. For 4C 09.17 A, the molecular outflow has the higher energetics. In 4C 09.17 B, we do not detect an ionized outflow, likely due to obscuration effects since we measure a high concentration of molecular gas with a high $\rm A_{V}$ value near the center of galaxy and 4C 09.17 B galaxy also has a red $V-K$ color of $>$5.35. Hence it is unclear how the energetics are distributed between the ionized and molecular gas phase of the outflow. For most of our objects, the distribution in energetics is consistent with recent theoretical predictions by \citet{Richings20}. The coupling efficiency between the multi-phase outflow and the bolometric luminosity of the quasar is consistent with theoretical predictions by \citet{Hopkins10,Choi12,Costa18}, where a minimum of 0.1-0.5$\%$ of the quasar's bolometric luminosity is expected to transfer into the kinetic luminosity of the outflow for there to impact star formation processes.

\subsection{Missing mass in galactic outflows}

We do not have any measurement of the neutral atomic gas mass in the outflow. Yet the neutral atomic phase likely exists within each outflow since we observe ionized emission through the optical \oiii and \ha lines. Theoretical work by \citet{Dempsey18} suggests that in the absence of a substantial amount of neutral gas, the \oiii transition would not be present, and most of the gas would be over-ionized (e.g., into [O IV]). At the same time, theoretical work by \citet{Richings18} has shown that a substantial amount of the molecular gas in an outflow is expected to be in the warmer molecular gas phase, with temperatures on the order of 400 K. The CO transition that trace the molecular gas reservoir within our study are only capable of tracing the cold molecular gas phase at temperatures of 40-70 K. The actual total momentum flux ratios are still likely lower limits, so energy-conserving outflows driven through a quasar mechanism are the most likely scenario for explaining the driving mechanism behind these galactic-scale outflows. Future observations with the MIRI instrument aboard \textit{JWST} will trace the warm molecular gas phase through the rest-frame mid-infrared rotational transitions of hydrogen. Observations with future 30-meter class telescopes will be able to probe the neutral gas phase through the Na D, [OI] 6300 \AA\ lines. Furthermore, the Square Kilometer Array (SKA) will enable us to probe the neutral gas phase directly through the 21 cm hydrogen line. 

\subsection{Molecular gas depletion time scales}

This section we explore the molecular gas depletion time scale in the galaxies within our sample. The molecular gas depletion time scale is defined as $t_{depletion, SFR} = M_{molecular}/SFR$. We will also compare the star formation depletion scale to the outflow depletion scale defined in a similar manner; $t_{depletion, outflow} = M_{molecular}/\dot{M}_{outflow}$. We compare the depletion time scale of the molecular gas due to star formation and from both the ionized and the molecular gas outflows. In all cases, we find a molecular gas depletion time scale of $<$ 31 Myr, with $t_{depletion, outflow}$ having the shorter timescale. The infrared-derived star formation rates are two orders of magnitude too high to be supported by the current molecular surface gas density, suggesting that the IR emission is likely contaminated by the quasar. While the maximum derived star formation rate from the total-infrared emission can be comparable to the multi-phase gas outflow rates, at present time, star formation rates expected from the Kennicutt-Schmidt (KS) law are two orders of magnitude smaller. The gas depletion timescale due to the KS-derived star formation rates is about two orders of magnitude higher than the depletion time scale due to outflows. Due to the expected low star formation rate from the low molecular gas surface density, in the next few Myr, the depletion time scale will be dominated by the galactic scale outflows. The depletion time scales from each source for the systems in our survey are presented in Table \ref{tab:depletion}. Not only are we catching these systems when a substantial fraction of the gas is in an outflow state, the rate of molecular gas depletion is dominated by the quasar driven outflows.

\begin{deluxetable*}{ccccc}
\tablecaption{Measured depletion time scales based on the star formation activity and multi-phase outflows in our sample. t$_{depletion,SFR}$ is the depletion time scale of the current molecular reservoir using the highest star formation rate observed. t$_{depletion,outflow}$ is the depletion time scale due to the outflows, and t$_{depletion,KS}$ is the depletion time scale based on the KS law for the observed molecular gas surface density. t$_{depletion,MS}$ is the expected depletion time scale for a galaxy at the measured stellar mass on the galaxy main sequence. \label{tab:depletion}}

\tablehead{
\colhead{Source}&
\colhead{t$_{depletion,SFR}$}&
\colhead{t$_{depletion,outflow}$}&
\colhead{t$_{depletion,KS}$} &
\colhead{t$_{depletion,MS}$}\\
\colhead{}&
\colhead{Myr}&
\colhead{Myr}&
\colhead{Myr}&
\colhead{Myr}}
\startdata
4C 09.17 A RL &  & 6$\pm$1 & 500$\pm$50 & 700 \\
4C 09.17 B RQ &  & 4$\pm$1 & 178$\pm$18 & 700 \\
7C 1354 & 10$\pm$4 & 6$\pm$3 & 300$\pm$100& 700 \\
3C 298 & 7$\pm$1 & 4$\pm$1 & 275$\pm$40 & 700 \\
3C 318 & $<$ 1 & $<$ 1.7$\pm$0.6 & & 800 \\
4C 22.44 & $<$ 31$\pm$3 & $<$3$\pm$2 & & 800 \\
4C 05.84 & $<$1.5 & $<$0.8$\pm$0.5 & & 700 \\
\enddata
\end{deluxetable*}

In the last several years, there have been many molecular gas observations in massive galaxies near cosmic noon. We can compare the depletion time scales that we observe within our systems to those that do not have powerful quasars. We compare the expected depletion time scale due to star formation for an average galaxy in our stellar mass range to the observed depletion time scale due to the outflows. \citet{Tacconi18} measured the depletion time scale for galaxies at an extensive range of redshifts and stellar masses. The dynamical mass \citep{Vayner19b} of the galaxies within our sample range from $10^{10.5-11.5}$ \msun, with star formation rates of 30-1330 \myr. This places them on the massive end of the galaxy luminosity function and the galaxy main sequence (MS) at $z=2$ that relates the star formation rate of a galaxy to its stellar mass. For galaxies with a mass in the range of $10^{10.5-11.5}$ \msun, the expected depletion time scale is around 700-800 Myr for galaxies on the star-formation main sequence at $z\sim2$. The range in the molecular gas depletion time scale for galaxies in the mass range of $10^{10.5-11.5}$ \msun\ is $0.3-3.5$ Gyr. Our systems appear to show a depletion time scale of the molecular reservoir that is least 100 times faster, indicating that powerful quasars can play a role in having shorter depletion times compared to star formation in a typical massive galaxy. While the \ha derived star formation rates place the sample quasar hosts on or below the star forming main sequence at z$\sim$2, the much larger FIR-derived star formation rates place them well above the MS.  We therefore estimate a range of depletion times in Table \ref{tab:depletion} covering the range of estimated star formation rates for each source.

The rapid depletion time scales of the cold molecular gas reservoir and the present amount of molecular gas has a profound implication for the evolution of these galaxies from cosmic noon to the present day. From the Keck/OSIRIS observations, we learned that these galaxies are offset from the local \msigma and \mstellar relationships \citep{Vayner19b}, indicating that the galaxies are under-massive for the mass of their SMBH. We estimated that they require a continuous stellar mass growth of approximately 100 \myr\ between z=2 and z=0 to land onto the local scaling relations. For each system, we have computed the molecular gas at the systemic redshift of the quasar within the radius where the dynamical masses are calculated in \cite{Vayner19b}. We find molecular gas mass in the range of $0.3-10\times10^{9}$ \msun, while in systems without detection of CO emission at the systemic redshift, we have placed limits of $<1\times10^{9}$ \msun. None of the systems have the molecular gas mass necessary to increase their stellar mass to bring them close to the local scaling relations between the SMBH mass and the stellar mass of the galaxy.

Furthermore, the majority of the molecular gas is in the galactic outflows rather than in the host galaxy; hence the overall stellar mass is still not going to increase by a substantial amount through star-formation. This study further indicates that a large new fresh reservoir of gas is necessary to accrete from the circumgalactic medium to replenish the fuel necessary for future star formation, and/or a large number of mergers is expected between z=2 and z=0 \citep{Burke13,Cooke19}. Our results further indicate that strong quasar feedback occurs before galaxies assemble onto the local scaling relations.

\section{Conclusions}\label{sec:conc}

We have conducted a study of the molecular gas properties in 6 radio-loud quasar host galaxies at $1.4<z<2.3$ through ALMA band 4 observation of the CO (3-2) and CO (4-3) rotational transitions at sub-arcsecond resolution. The survey aimed to study the molecular gas morphologies and kinematics to address the energetics and evolution of these massive galaxies. We resolve the molecular gas reservoirs in 5 galaxies and detect molecular outflows in 4 systems. Coupling these observations with high spatial resolution observations with Keck/OSIRIS we are able to study the ionized gas and molecular gas to gauge the impact of quasar activity on the star formation properties of the host galaxies.

\begin{itemize}
    \item We detect CO emission on kpc scale in 5/6 quasar host galaxies.
    \item We detect high-velocity clumps in 3C 318, 4C 09.17A, and 4C 05.84, offset up to -1200 km/s while in 3C 298 and 4C 09.17B we detect broad CO emission lines. We interpret these spectral features as molecular outflows. We derive outflow rates of 168-2500 \myr, combining the energetics of the ionized to the molecular gas outflows, we find that the dominant outflow mechanism is through an energy-conserving shock from the quasar jets.
    \item We find that the outflowing gas to be predominantly in the molecular gas phase while the momentum flux and kinetic luminosity resides in the ionized gas phase for 3/4 multi gas-phase outflows.
    \item While the maximum star-formation rate from total infrared luminosity show a similar depletion time scales to the outflows, due to a much lower molecular gas surface densities than expected, the outflow depletion time scales are at least 50 times faster. The depletion time scale compared to the expected value for an average galaxy on the galaxy star formation main sequence at the stellar mass of our survey is about 100 times faster.
    \item A substantial accretion of fresh gas from the circumgalactic medium, and/or stellar growth through dry mergers is necessary for these galaxies to achieve the expected mass based on their central SMBH mass using local scaling relations. The current molecular gas reservoir is insufficient to provide a significant amount of fuel for star formation.
    \item In the 4C 09.17 system, we find a molecular gas outflow in its quasar host galaxy, as well as a nearby dust-obscured galaxy, indicating evidence for feedback occurring well before the coalescence phase of this merger system.
\end{itemize}

\bibliography{bib}{}
\bibliographystyle{aasjournal}

\acknowledgments
The authors wish to thanks Jim Lyke, Randy Campbell, and other SAs with their assistance at the telescope to acquire the Keck OSIRIS data sets. We would like to thank Erica Keller, Melissa Hoffman, and Loreto Barcos Munoz for assistance with ALMA data reduction and imaging at NRAO. We want to thank the anonymous referee for their constructive comments that helped improve the manuscript. This paper makes use of the following ALMA data: ADS/JAO.ALMA 2013.1.01359.S, ADS/JAO.ALMA 2017.1.01527.S. ALMA is a partnership of ESO (representing its member states), NSF (USA) and NINS (Japan), together with NRC (Canada), MOST and ASIAA (Taiwan), and KASI (Republic of Korea), in cooperation with the Republic of Chile. The Joint ALMA Observatory is operated by ESO, AUI/NRAO and NAOJ. The National Radio Astronomy Observatory is a facility of the National Science Foundation operated under cooperative agreement by Associated Universities, Inc. The data presented herein were obtained at the W.M. Keck Observatory, which is operated as a scientific partnership among the California Institute of Technology, the University of California and the National Aeronautics and Space Administration. The Observatory was made possible by the generous financial support of the W.M. Keck Foundation. The authors wish to recognize and acknowledge the very significant cultural role and reverence that the summit of Maunakea has always had within the indigenous Hawaiian community. We are most fortunate to have the opportunity to conduct observations from this mountain. This research has made use of the NASA/IPAC Extragalactic Database (NED) which is operated by the Jet Propulsion Laboratory, California Institute of Technology, under contract with the National Aeronautics and Space Administration.

\software{OSIRIS DRP \citep{OSIRIS_DRP}, CASA\citep{McMullin07}, Matplotlib \citep{Hunter07}, SciPy \citep{2020SciPy-NMeth}, NumPy \citep{2020NumPy-Array}, Astropy \citep{Astropy18}}

\end{document}